\documentclass[%
reprint,
superscriptaddress,
showpacs,showkeys,
amsmath,amssymb,
aip,
jvsta,
]{revtex4-1}

\usepackage{graphicx,subfigure,float}
\graphicspath{{Figures/}}
\usepackage{adjustbox}
\usepackage{dcolumn}
\usepackage{bm}
\usepackage{color}
\usepackage{hyperref}

\begin{document}


\title{The role of ionization fraction on the surface roughness, density and interface mixing of the films deposited by thermal evaporation, dc magnetron sputtering and HiPIMS: An atomistic simulation}
\author{Movaffaq Kateb}
\email[Corresponding author email address: ]{mkk4@hi.is}
\affiliation{Science Institute, University of Iceland,
Dunhaga 3, IS-107 Reykjavik, Iceland}

\author{Hamidreza Hajihoseini}
\affiliation{Science Institute, University of Iceland,
Dunhaga 3, IS-107 Reykjavik, Iceland}
\affiliation{Department of Space and Plasma Physics, School of Electrical Engineering and Computer Science, KTH Royal Institute of Technology, SE-100 44, Stockholm, Sweden}

\author{Jon Tomas Gudmundsson}
\affiliation{Science Institute, University of Iceland,
Dunhaga 3, IS-107 Reykjavik, Iceland}
\affiliation{Department of Space and Plasma Physics, School of Electrical Engineering and Computer Science, KTH Royal Institute of Technology, SE-100 44, Stockholm, Sweden}

\author{Snorri Ingvarsson}
\affiliation{Science Institute, University of Iceland,
Dunhaga 3, IS-107 Reykjavik, Iceland}

\date{\today}

\begin{abstract}
We explore the effect of ionization fraction on the epitaxial growth of Cu film on Cu (111) substrate at room temperature. We compare thermal evaporation, dc magnetron sputtering (dcMS) and high power impulse magnetron sputtering (HiPIMS). Three deposition conditions i.e.\ fully neutral, 50~\% ionized and 100~\% ionized flux were considered as thermal evaporation, dcMS and HiPIMS, respectively, for $\sim$20000 adatoms. It is shown that higher ionization fraction of the deposition flux leads to smoother surfaces by two major mechanisms i.e.\ decreasing clustering in the vapor phase and bi-collision of high energy ions at the film surface. The bi-collision event consists of local amorphization which fills the gaps between islands followed by crystallization due to secondary collisions. We found bi-collision events to be very important to prevent island growth to become dominant and increase the surface roughness. Regardless of the deposition method, epitaxial Cu thin films suffer from stacking fault areas (twin boundaries) in agreement with recent experimental results. In addition, HiPIMS deposition presents considerable interface mixing while it is negligible in thermal evaporation and dcMS deposition, those present less adhesion accordingly. 

\end{abstract}


\maketitle


\section{Introduction}
High power impulse magnetron sputtering (HiPIMS) is an ionized physical vapor deposition (PVD) technique that has attracted significant interest over the past two decades.\citep{helmersson06:1,gudmundsson12:030801} By pulsing the cathode target to a high power density with unipolar voltage pulses, at low duty cycle, and low repetition frequency high electron density is achieved. \citep{helmersson06:1,gudmundsson12:030801,gudmundsson10:1360} In conventional dc magnetron sputtering (dcMS), the plasma density is limited by the thermal load on the target, and is usually on the order of 10$^{15}-10^{17}$~per cubic meter \citep{seo04:1310,seo06:256,sigurjonsson08:062018} and the ionization mean free path for the sputtered material is of the order of 50~cm. \citep{gudmundsson10:1360} Thus the fraction of ionized species of the target material is therefore low, typically well below 10~\%. \citep{christou00:2897} Consequently, the majority of particles reaching the substrate surface are electrically neutral and the ions are ions of the rare working gas. In HiPIMS, this problem is solved by applying high power impulses with a low duty cycle. The high power leads to peak electron densities exceeding 10$^{19}$~m$^{-3}$ in the vicinity of the cathode target. \citep{gudmundsson02:249,bohlmark05:346,meier18:035006} The high density of electrons increases the probability for ionizing collisions between the sputtered atoms and energetic electrons, and results in a high degree of ionization of the sputtered material. Values up to 70\% have been reported for the ionization flux fraction in the case of Cu \citep{kouznetsov99:290} and copper ions have been observed to be dominant (up to 92\%) in total ion fluxes to the substrate. \citep{vlcek07:45002} Beside the atoms and ions that collide with the substrate have energy distribution ranging 0 -- 100~eV which is higher than that of dcMS deposition (0 -- 40~eV). \citep{bohlmark06:1522,lin09:3676,vlcek07:45002} As a result HiPIMS presents denser, \citep{samuelsson10:591} smoother \citep{magnus11:1621,sarakinos07:2108} and void-free \citep{alami05:278} coatings compared to conventional sputtering methods. 
In spite of huge theoretical and experimental efforts on understanding different aspects of HiPIMS deposition, the atomistic mechanisms that contribute to the film properties are not well demonstrated so far. 

Atomistic simulations, namely Monte Carlo (MC) \citep{muller1985,muller1986,muller1986jap,dodson1990} and molecular dynamics (MD) \citep{muller1987prb,muller1987} have shown promise in investigation of PVD processes owing to their atomistic resolution. In this regard, PVD in the absence of ions and vapor phase collisions has been extensively studied. However, most of these simulations only cover low energy PVD, similar to molecular beam epitaxy, where the evaporated species have energy in the 0.1 -- 2~eV range. \citep{zhou1997} The films deposited at such conditions and at relatively low temperatures are mainly suffering from porous and columnar microstructure \citep{muller1985,henderson1974,kim1977} which is more pronounced in oblique deposition. \citep{henderson1974,kim1977,zhou1997,hubartt2013} While increased substrate temperature \citep{schneider1985,schneider1987prb,smith1996,zhou1997,zhang1998} and/or increased adatom energy \citep{smith1996,zhou1997,zhang1998} leads to a void-free film. This is mainly due to the fact that low energy deposition encourages island growth but the growth mode changes to layer-by-layer (Frank -- van der Merwe) growth as the incident atom energy is increased to 10~eV. \citep{gilmore1991,zhang1998} 
This higher energy 10 -- 40~eV causes interruption of layer-by-layer growth and leads to interface mixing between film and substrate. \citep{sprague1996,zhang1998,lugscheider1999} Since the interface mixing has some similarities to the thermal spike in bulk ion mixing, energetic deposition is considered as simplified model of sputter deposition in MD simulation. \citep{sprague1996} For instance, it has been shown that pollution of sputtered flux with high energy atoms, as mimic of partially ionization flux, leads to amorphization of the film \citep{houska2014} and fully energetic deposition gives smooth amorphous film. \citep{chen2015}

An alternative method to model sputtering conditions is demonstrated in atomistic simulation of ion assisted PVD. \citet{muller1986,muller1986jap,muller1987prb,muller1987} was probably the first who considered a deposition using a flux consisting of both neutral adatoms and rare ions. He showed that bombarding the film with low energy rare ions removes bridging on top of the voids and thus leads to densification and texture refinement. \citep{muller1987prb,muller1987} He studied the effect of rare ion to neutral ratio, the rare ion energy and adatom energy on the void formation in the film which can be correlated to the tensile stress in the film. It has also been shown that ion-assisted PVD can cause texture refinement. \citep{dong1998,dong1999} In addition, ion-assisted deposition can be used for more uniform deposition of Cu into trenches and vertical interconnect access (VIA). \citep{hwang2002,hwang2003} Furthermore, it has been shown that for the case of Cu deposition, the ion energy has major effect on the surface roughness compared to ion incident angle. \citep{su2006} More recently, \citet{xie2014} proposed a distribution function to mimic the kinetics energies of sputtered flux at the substrate surface in MD simulation. This allows a more realistic simulation but the method is still limited to a distribution function, e.g. Thompson.  

In spite of these huge efforts, many of the above mentioned studies suffer from being performed in 2D, \citep{muller1987prb,muller1987} using simplified force fields, e.g.\ hard sphere or LJ, \citep{henderson1974,kim1977,dong1998,lugscheider1999} and limited number of deposited species. \citep{lugscheider1999} Thus, the previous studies were limited to only early stage of deposition, due to lack of computation power. There are also some studies on the accelerated simulation that are focused on the more realistic (slow) deposition rates. \citep{sprague2002,hubartt2013} The energy distribution in the flux also has been neglected which might be reasonable assumption in thermal evaporation but it is necessary for realization of ionized PVDs. \citep{bohlmark06:1522} In addition the effect of ionized flux on the film microstructure has never been discussed. Thus, they were unable to reflect ion-ion repulsion within the plasma as well as resputtering of the film due to bombardment of high energy ions.

The aim of the present study is to consider the effect of ionized flux of the deposition species as a major difference between evaporation, dcMS and HiPIMS deposition in the MD framework. To this end, the film density, surface roughness, microstructure and interface mixing are probed during film deposition at the atomic resolution.
\section{Method}

MD simulations were performed by solving Newton's equation of motion \citep{allen1989} using the large-scale atomistic/molecular massively parallel simulator (LAMMPS)  open source code. \citep{plimpton1995,plimpton2012} \footnote{LAMMPS website, \url{http://lammps.sandia.gov/}, distribution 14-April-2018}

The thermal evaporation flux, dcMS flux and the HiPIMS flux were assumed to be fully neutral, 50~\% ionized and fully ionized, respectively. The solid phase and neutrals interaction was modeled using embedded-atom method (EAM) potentials. \citep{daw1983,daw1984} The total potential energy of atom $i$, $E_i$ is described by
\begin{equation}
	E_i=F_i(\rho_i)+\frac{1}{2}\sum_{i\neq j}\phi_{ij}(r_{ij})
\end{equation}
where $F_i$ is the embedding energy of atom $i$ into electron density $\rho_i$ and $\phi_{ij}$ is a pair potential interaction of atom $i$ and $j$ at distance $r_{ij}$. 

The multi-body nature of the EAM potential is a result of the embedding energy term i.e.\ $\rho_i$ itself depends on electron density of neighboring atoms $\rho_{ij}$
\begin{equation}
	\rho_{i}=\sum_{i\neq j}\rho_{ij}(r_{ij})
\end{equation}

The ion-ion interaction in the flux was modeled via Ziegler-Biersack-Littmark (ZBL) \citep[chap.~2]{ziegler1985} potential which takes into account both short range Coulombic interaction and long range screening.
\begin{equation}
	V(r_{ij})=\frac{Z_iZ_je^2}{4\pi\varepsilon_0r_{ij}}\Phi\left(\frac{r_{ij}}{a}\right)
\end{equation}
where the $Z_i$ and $Z_j$ are the atomic numbers of ion $i$ and ion $j$ the ions that belong to Coulombic term. $e$ and $\varepsilon_0$ stand for elementary charge and vacuum permitivity, respectively.

The universal screening function in reduced unit is defined 
\begin{equation}
	\Phi\left(\frac{r_{ij}}{a}\right)=\sum_{n=1}^4a_n e^{-c_nr_{ij}/a}
\end{equation}
where $a$ is the ZBL modification of Bohr's universal reduced coordinate with 0.8853 derived from Thomas-Fermi atom
\begin{equation}
	a=\frac{0.8853a_0}{Z_i^{0.23}+Z_j^{0.23}}
\end{equation}
with $a_0$ being Bohr radius and $a_n$ is normalizing factor i.e. $\sum a_n=1$.
\begin{eqnarray}
    &&a_n= 0.18175, 0.50986, 0.28022, 0.02817\nonumber\\
    &&c_n= 3.19980, 0.94229, 0.40290, 0.20162\nonumber
\end{eqnarray}

We would like to remark that the ZBL potential present 5~\% standard deviation from experimental values while the deviation for the popular Moliere potential can be very large (237~\%). \citep[chap.~2]{ziegler1985} The cut off was considered to be 2.552~{\AA} which is large enough to model sputtering\citep{kammara2016} and a switching function was also considered to smoothly ramp energy and force to zero at cutoff.

Ion-neutral and ion-film interactions were modeled using a hybrid based on both EAM and ZBL potentials. This allows resputtering from the film due to the repulsive force of the ZBL potential. Once an ion collides with the surface it may be either scattered back or it stabilizes at the surface. If it stands at the surface for a short time (1~ps) or implants into sublayers, it enters into the solid phase and thus its interatomic potential is defined by EAM afterwards. This may multiply the computation cost but it is necessary to realize and retain deposition condition otherwise surface etching and incident ions scattering are expected.

The substrate were considered to be a single crystal Cu with its $\langle111\rangle$ orientation parallel to the growth direction, which means a (111) plane is exposed to the deposition flux. The initial configuration consisted of a fixed monolayer, a thermostat layer (3 monolayers) and a surface layer (12 monolayers) all with 77$\times$90~{\AA}$^2$ lateral dimensions. The initial velocities of substrate atoms were defined randomly from a Gaussian distribution at the appropriate temperature of 300~K and the substrate energy was minimized afterwards. 

For all deposition methods, the flux ratios atoms/ions were inserted 150 nm above the substrate surface with random energy ranging 0 -- 100~eV. We assumed a uniform distribution for all three deposition methods. In the case of dcMS, 50\% ionization is expected to have the same energy distribution for ions and neutrals. The inserting process was a single atom/ion each 0.1~ps with initial velocity parallel to the substrate normal which gave a linear equal deposition rate in all cases.The HiPIMS deposition is normally performed using 50 -- 400 $\mu$s long pulses \citep{gudmundsson12:030801} which is longer than the simulation times achieved in MD. Here the impulse nature of HiPIMS was neglected and deposition was assumed to remain for the entire time. 

The time integration of the equation of the motion was performed regarding microcanonical ensemble (NVE) with a timestep of 5~fs. Since practical deposition is performed in the vacuum, the heat associated with particle's collision cannot be removed so efficiently and hence, the NVE ensemble  provides a realistic representation of such systems. The Langevin thermostat \citep{schneider1978} was only applied to the specified layer with a damping of 5~ps. It is worth mentioning that the damping is not due to the fact that Langevin thermostat does the time integration. But it modifies the forces instead which reproduce deceleration of ions implanted into film with unique precision.

The first and simplest structure analysis is offered by $G(r)$ or pair correlation function written as
\begin{equation}
    G(r)=\langle4\pi r^2\rho_adr\rangle_T
\end{equation}
where $\rho_a$ is the atom numbers density, $r$ is the distance from reference particle and $dr$ determines the bin size. The angle brackets i.e. $\langle\rangle_T$ denote time average at constant $T$. 

The $G(r)$ describes how density varies as a function of distance in a system of particles from a reference particle. This results in a pattern of several peaks corresponding to number and distance of nearest neighbors (NNs) which applies to a wide range of materials. The amorphization as a result of ion bombardment causes variation in the density and can be detected by shifting and broadening of peaks in the $G(r)$ pattern. However complex solid-state transition such as fcc to hcp with constant coordination number and even distance, is very hard to determine with $G(r)$.

Common neighbor analysis (CNA) has shown to be a promising tool for structure characterization due to possibility of distinction between allotropic transitions and melting process. The CNA identifies crystal structure of each atom based on the concept of bond-orientational order parameter (BOP) developed by \citet{steinhardt1983}. Briefly, the CNA determines local crystal structure based by decomposition of 1st NNs obtained from $G(r)$ into different angles. \citep{tsuzuki2007} Unlike to the BOP, CNA is sensitive to angles between pairs of NNs and can distinguish between fcc and hcp. Thus a twin grain boundary can be determined based on slight angle difference between 1st NNs while it holds entire properties of an fcc atom.

The Ovito package\footnote{Ovito website, \url{http://ovito.org/}, Version 2.9.0} were used to generate atomistic illustrations. \citep{stukowski2009}

\section{Results and discussion}
\subsection{Interface mixing}

\begin{figure*}
    \centering
	\subfigure{\includegraphics[width=1\linewidth]{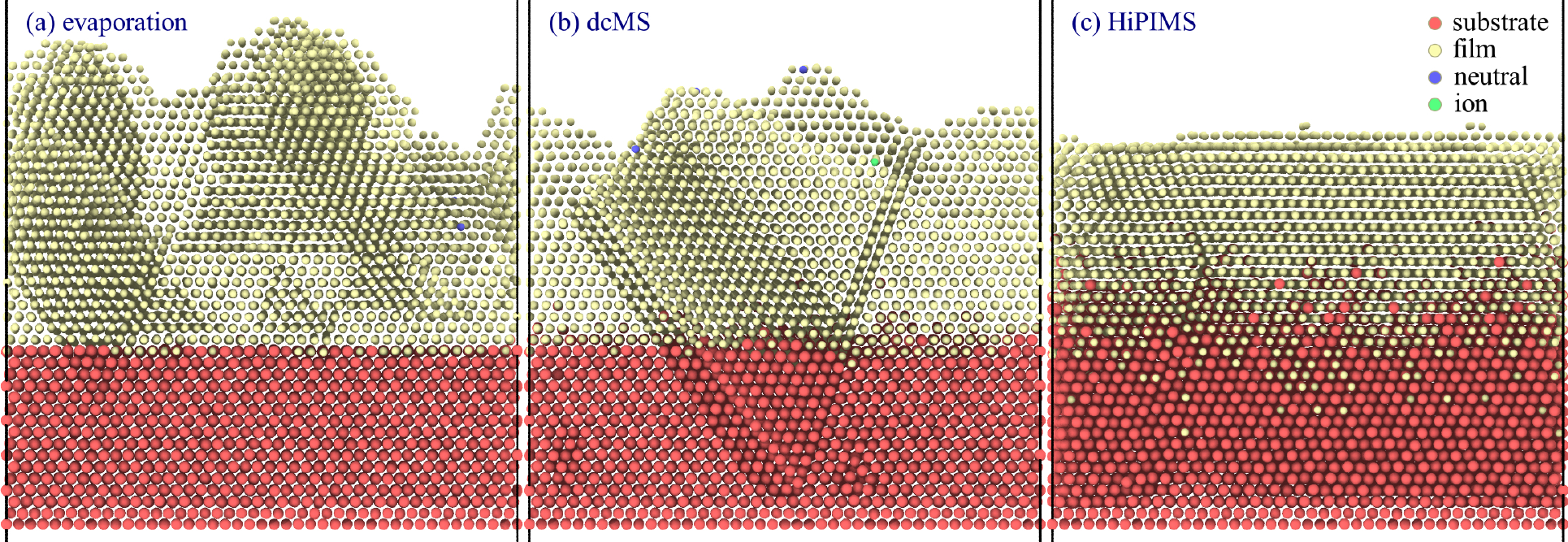}}
    \caption{Illustration of interface mixing using (a) thermal evaporation, (b) dcMS and (c) HiPIMS after 2.5~ns deposition. The red, green, blue and yellow are indicating substrate, neutral, ions and film atoms.}
    \label{fig:mix2.5ns}
\end{figure*}

Fig.~\ref{fig:mix2.5ns} shows the Cu films in yellow deposited by the three different methods on a identical flat substrate indicated by red. For thermal evaporation and dcMS deposition shown in Fig.~\ref{fig:mix2.5ns} (a) and (b) no interface mixing is observed. It can be seen that the full ionization of the depositing species in HiPIMS effectively increases the interface mixing (see Fig.~\ref{fig:mix2.5ns} (c)). Thus, it can be expected that HiPIMS deposited film present the highest adhesion to the substrate while thermal evaporation and dcMS present negligible difference in terms of adhesion. Moreover, better electrical contact can be expected due to interface mixing. It has been already shown using MC simulation that when an ion with a few hundreds of eV energy strikes the surface of a low density film, with less than 80~\% of theoretical density, it can penetrate to an average depth of a few nm. \citep{muller1986,muller1986jap} In the case of our HiPIMS deposition, adatoms can be found at maximum at 1.5~nm depth of substrate surface. This difference might be due to the fact that the previous MC simulation were performed in 2D, using Ar$^+$ ions and Moliere potential. The interface mixing has been found to be sensitive to the temporal lattice excitations localized in the vicinity of atom impacts. \citep{sprague1996} In the present result the interface mixing is associated with alternating localized amorphization and mixing due energetic impacts. In dcMS deposition, limited number of energetic impacts occurs at the surface and interface mixing is negligible while in HiPIMS deposition such event becomes dominant and thus interface mixing is considerable. These results are in agreement with the recent experimental comparison of Cu films deposited on Si with a native oxide using dcMS and HiPIMS. \citep{cemin2017} At identical conditions, only Cu deposited by HiPIMS can pass through the native oxide and form epitaxial film. 

\subsection{Surface roughness}
\begin{figure*}
    \centering
	\subfigure{\includegraphics[width=1\linewidth]{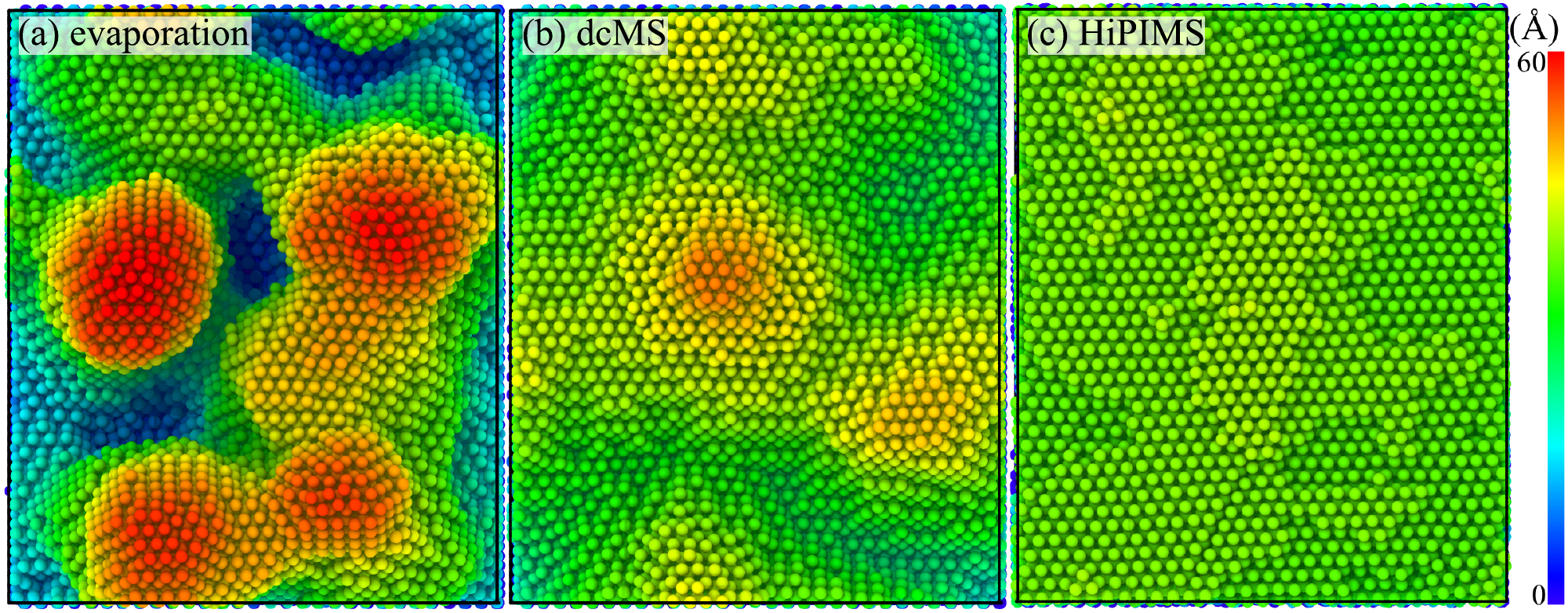}}
    \caption{The surface topology obtained using (a) thermal evaporation (b) dcMS and (b) HiPIMS deposition with similar deposition time and energy distribution. The deep blue indicates substrate surface and red denotes thickness higher than 6~nm.}
    \label{fig:Z2.5ns}
\end{figure*}

Fig.~\ref{fig:Z2.5ns} shows the top view of the films deposited by thermal evaporation, dcMS and HiPIMS with identical deposition time and energy distribution. The dark blue here shows the substrate surface and atoms that are 6~nm above the substrate are identified by red. It can be clearly seen that the thermal evaporated film presents very rough surface compared to the sputtered films. This is due to the fact that during thermal evaporation neutral atoms form clusters before arriving at film/substrate surface. One may think the surface roughness obtained in thermal evaporation here is an artificial effect of relatively high deposition rate or short simulation time compared to time required for surface diffusion. Such island growth has been reported for deposition of Cu on Cu with experimental rate and modeling diffusion process through accelerated dynamic simulation.\citep{hubartt2013} Thus, the film obtained by thermal evaporation is extremely non-uniform at the atomic level. In the HiPIMS deposition, however, the repulsion between ions does not allow clustering when maximum uniformity of deposition occurs as can be seen in Fig.~\ref{fig:Z2.5ns}(c). Due to distribution of energy in the flux, neutrals/ions with higher kinetic energy are able to diffuse longer at the surface than low energy adatoms. As a result formation of islands is still possible in the ionized flux case. The secondary mechanism here is energetic impacts of ions into subsurface atoms which causes local amorphization and fills the gaps between islands with atomically flat surface. The energetic ions themselves are the result of strong repulsive force between ions. Further collision of energetic ions cause recrystallization of amorphous regions which maintain smooth surface. We observed both of the above mentioned mechanisms i.e.\ clustering and energetic collision during dcMS deposition that give an intermediate surface roughness as seen in Fig.~\ref{fig:Z2.5ns}(b). 

Previously, it has been claimed that the only mechanism of redistribution of surface atoms is collapse of height advantaged islands at low energy deposition ($\sim$2~eV) and ballistic displacement of atoms at higher energies (in the 2 -- 10~eV range). \citep{gilmore1991} We did not observe such mechanisms even during thermal evaporation which gives columns with an average cross section $\sim$3~nm and 6~nm hight (cf. Fig.~\ref{fig:Z2.5ns}(a)). In fact, the previous study was limited to two or three monolayer islands and thus was able to reflect the early stage of deposition. The second difference here arises from the fact that we applied a distribution of energy and ionization fraction to the flux which leads to more realistic result compared to flux with mono-dispersed energy.

\subsection{Film density}

\begin{figure}
	\centering
    \includegraphics[width=1\linewidth]{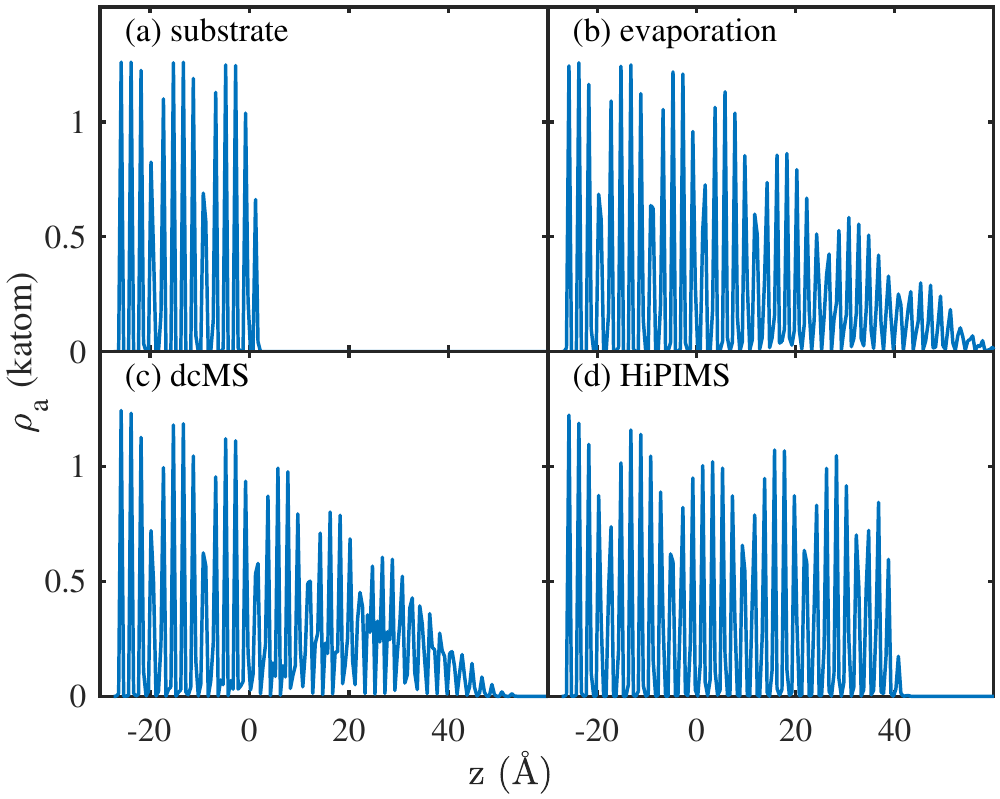}
    \caption{Histogram of spatial distribution of atoms (atomic density, $\rho_a$) along deposition direction with $z=0$ being the substrate surface.}
    \label{fig:rho}
\end{figure}

In Fig.~\ref{fig:rho} the atomic density, $\rho_a$, is compared along the deposition direction, $z$, for the three deposition methods. The substrate pattern after relaxation shows very sharp transition at the surface $z=0$ as seen in Fig.~\ref{fig:rho}(a). A similar pattern is obtained after HiPIMS deposition which is known as a sign of layer-by-layer growth \citep{muller1987prb} as can be seen in Fig.~\ref{fig:rho}(d). On the other hand thermal evaporation and dcMS deposition result in a gradual decay which is a characteristic of island growth \citep{gilmore1991} as can be seen in Fig.~\ref{fig:rho}(b) and (c), respectively.

\subsection{Temperature}

\begin{figure}
	\centering
    \includegraphics[width=1\linewidth]{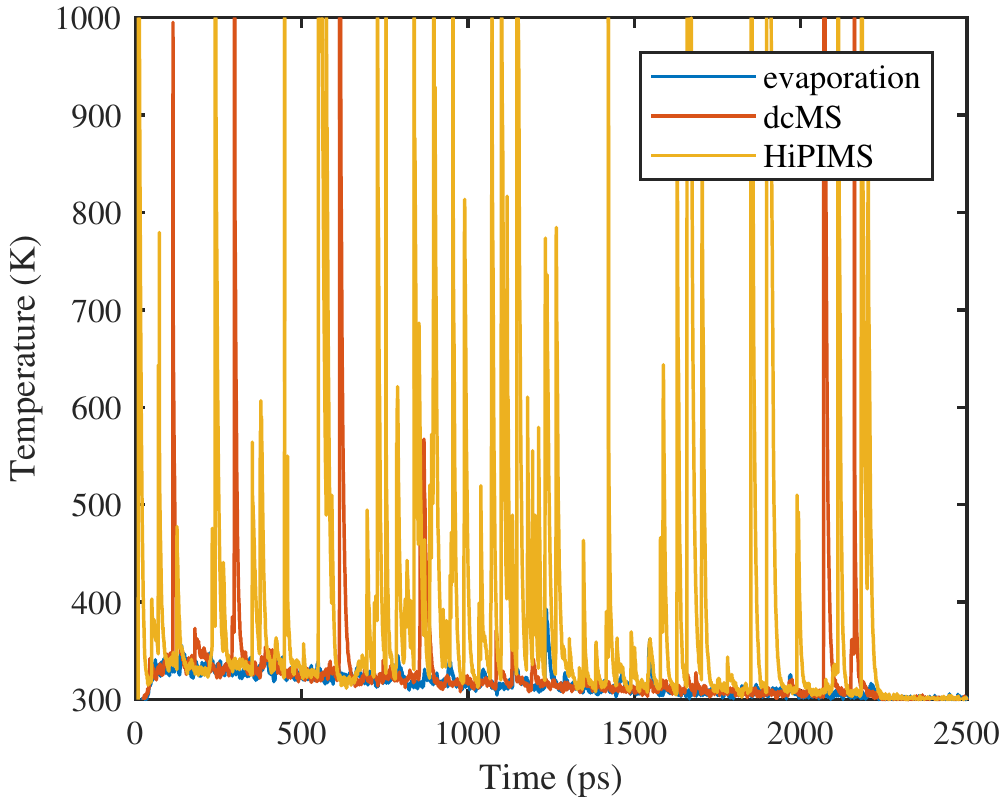}
    \caption{Variation of temperature in the thermostat layer during deposition using thermal evaporation, dcMS and HiPIMS}
    \label{fig:Tspike}
\end{figure}

Fig.~\ref{fig:Tspike} compares the variation of temperature with deposition time in thermal evaporation, dcMS and HiPIMS deposition. It can be seen that during thermal evaporation the temperature rises to $\sim$340~K within the early stage of deposition and gradually decays to 300~K during the deposition. We observe local peaks in thermal evaporation that belong to clusters colliding to the substrate surface. However the temperature variation related to cluster collision is very limited, ranging 10 -- 20~K. In contrast, the sputtering methods consisted of several thermal spikes some of them exceeding 1000~K. During deposition of atoms with energy in the range 0.1 -- 10~eV, without ions, the thermal spikes are not sufficiently strong to cause redistribution of surface atoms. \citep{gilmore1991} \citet{muller1986} showed by theoretical calculation that low energy ion impact can generate thermal spike and cause structure modification although, he used energetic Ar$^+$ ions with energy of 150~eV for demonstration of the effect. We did not observe any rearrangement at the surface due to small thermal peaks following cluster impacts. As mentioned before, we have noticed that the effect of thermal spikes is not only limited to the microstructure modification but also it is responsible for lower surface roughness obtained with the sputtering methods compared to the thermal evaporation. Since during HiPIMS deposition more thermal spikes occur than during dcMS deposition, it is expected to present a smoother surface (cf. Fig.~\ref{fig:Z2.5ns}) accordingly. This has indeed been observed experimentally. \citep{magnus11:1621,sarakinos07:2108}

\subsection{Microstructure}

\begin{figure*}
    \centering
	\subfigure{\includegraphics[width=1\linewidth]{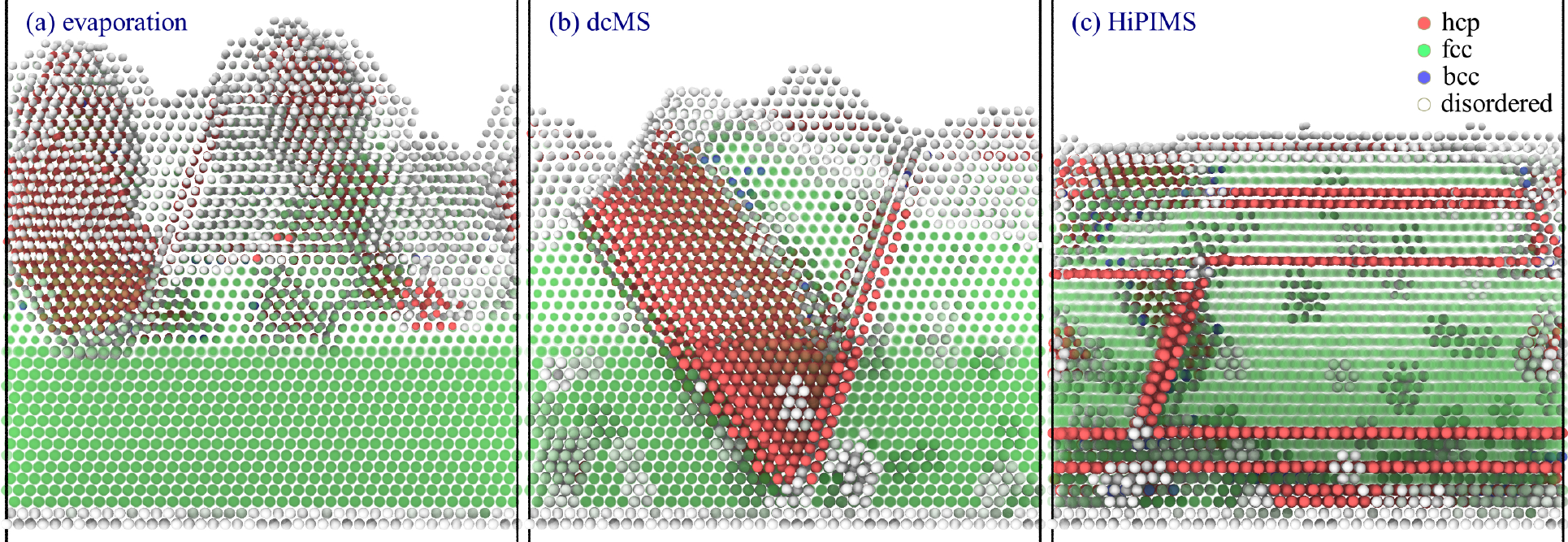}}
    \caption{Analysis of local structure using CNA with red, green, and white respectively being fcc, hcp and disordered atoms for (a) thermal evaporation, (b) dcMS and (c) HiPIMS. To distinguish between film/substrate, the film atoms are illustrated with smaller diameter.}
    \label{fig:micro}
\end{figure*}

The microstructures obtained by the three different deposition methods are shown in Fig.~\ref{fig:micro}(a) -- (c). The color contrast obtained by adoptive CNA which can distinguish between different crystal structures i.e.\ fcc, hcp, bcc and disordered atoms indicated by green, red, blue and white, respectively. In the current simulation dimensions, all methods providing single crystal Cu film aside from stacking faults (twin boundaries) and point defects. The formation of stable twin boundaries in the oblique deposition Cu on Cu (001) has been reported previously using accelerated MD simulation. \citep{hubartt2013} The existence of stacking fault areas has also been verified experimentally by polar mapping of the (111) planes in the epitaxial Cu deposited by thermal evaporation \citep{chen2013} and HiPIMS. \citep{cemin2017} Also we have recently demonstrated experimentally the existence of twin boundaries in epitaxial Ni$_{80}$Fe$_{20}$ (at.~\%) film deposited with both dcMS and HiPIMS. \citep{kateb2018} Temporal formation of stacking faults and twin boundaries in the plane of Cu (111) during sputtering deposition was observed which has been reported previously during low energy deposition of Cu on Cu (111) \citep{zhou1999} and Al on Cu (111). \citep{cao2010} 

It is worth noting that during thermal evaporation the substrate (indicated by bigger atoms) remains unchanged whereas in the dcMS and HiPIMS deposition both stacking faults and point defects are introduced into the substrate. This essentially means utilizing ions in the deposition flux enables modification of substrate structure in agreement with previous studies. \citep{muller1987prb,muller1987,dong1998} However, in these studies the ions were considered to be Ar$^+$ whose impact gives smaller momentum than Cu$^+$ ions utilized here. Thus in the previous studies the structure modification were limited to densification \citep{muller1987prb,muller1987} and reorientation of grains. \citep{dong1998}

\subsection{High energy collisions}

\begin{figure}
	\centering
    \includegraphics[width=1\linewidth]{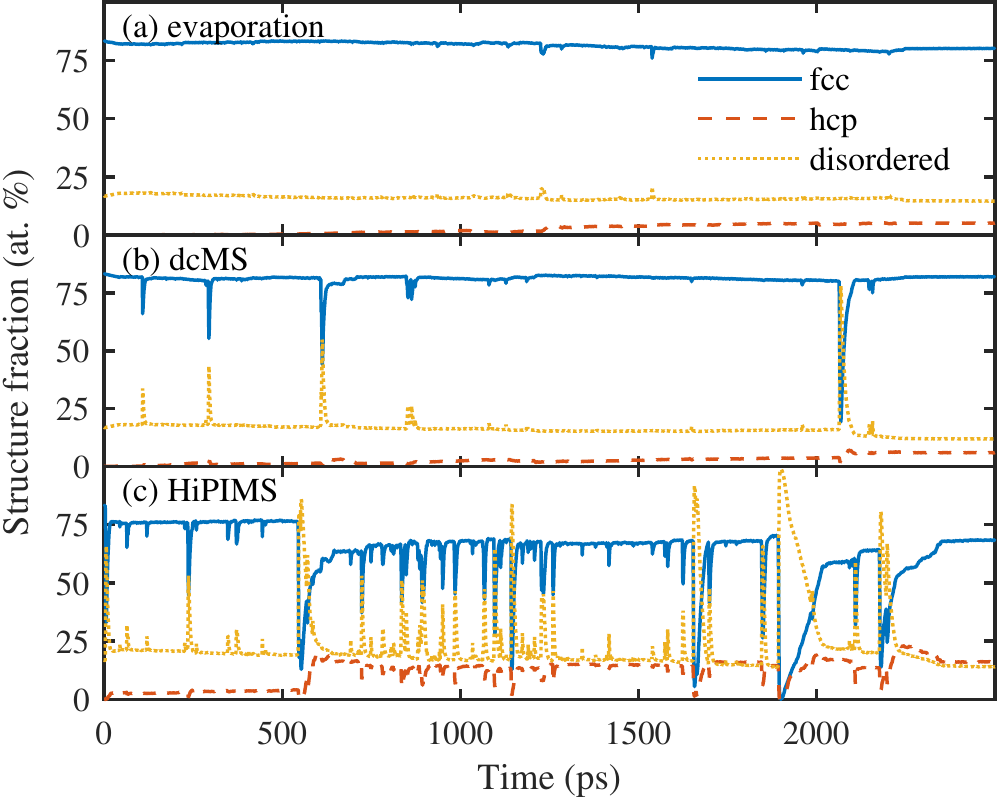}
    \caption{Variation of fcc, hcp and disordered fraction during deposition by (a) thermal evaporation, (b) dcMS and (c) HiPIMS.}
    \label{fig:CNA}
\end{figure}

Fig.~\ref{fig:CNA} shows the variation of structure fraction during deposition by each method. At the early stage of deposition, the largest fraction is the fcc structure due to single crystal substrate and minor fraction consists of disordered atoms, those located at the surface. During thermal evaporation, as shown in Fig.~\ref{fig:CNA}(a), these fractions are nearly constant except a slight increase in the fraction of hcp atoms which is associated with twin boundaries in the film (cf.\ Fig.~\ref{fig:micro}(a)). There are also some minor peaks in the fraction of disordered atoms and those are attributed to cluster impacts on the surface which generates temporary a amorphous phase at cluster-film interface. The dcMS deposition also presents similar result except for the fact that the peaks in the fraction of disordered atoms become considerable as shown in Fig.~\ref{fig:CNA}(b). It is worth mentioning that the peaks observed here are due to impacts of high energy ions rather than clusters.

In contrast with thermal evaporation and dcMS, the initial fractions are not conserved during HiPIMS deposition as shown in Fig.~\ref{fig:CNA}(c). For instance, $\sim$20~\% increase in hcp fraction is observed after a  significant amorphization peak at 550~ps which is associated with $\sim$20~\% decrease in the fcc fraction. Unlike both thermal evaporation and dcMS, the peaks in the fraction of disordered atoms are associated with pits in both fcc and hcp fraction. This is due to the fact that the fraction of hcp atoms or stacking fault areas generated during HiPIMS deposition is much larger ($\sim$20~\%) than for other methods. Thus the hcp fraction can be affected by high energy ion bombardment.

\begin{figure}
    \centering
	\subfigure{\includegraphics[width=1\linewidth]{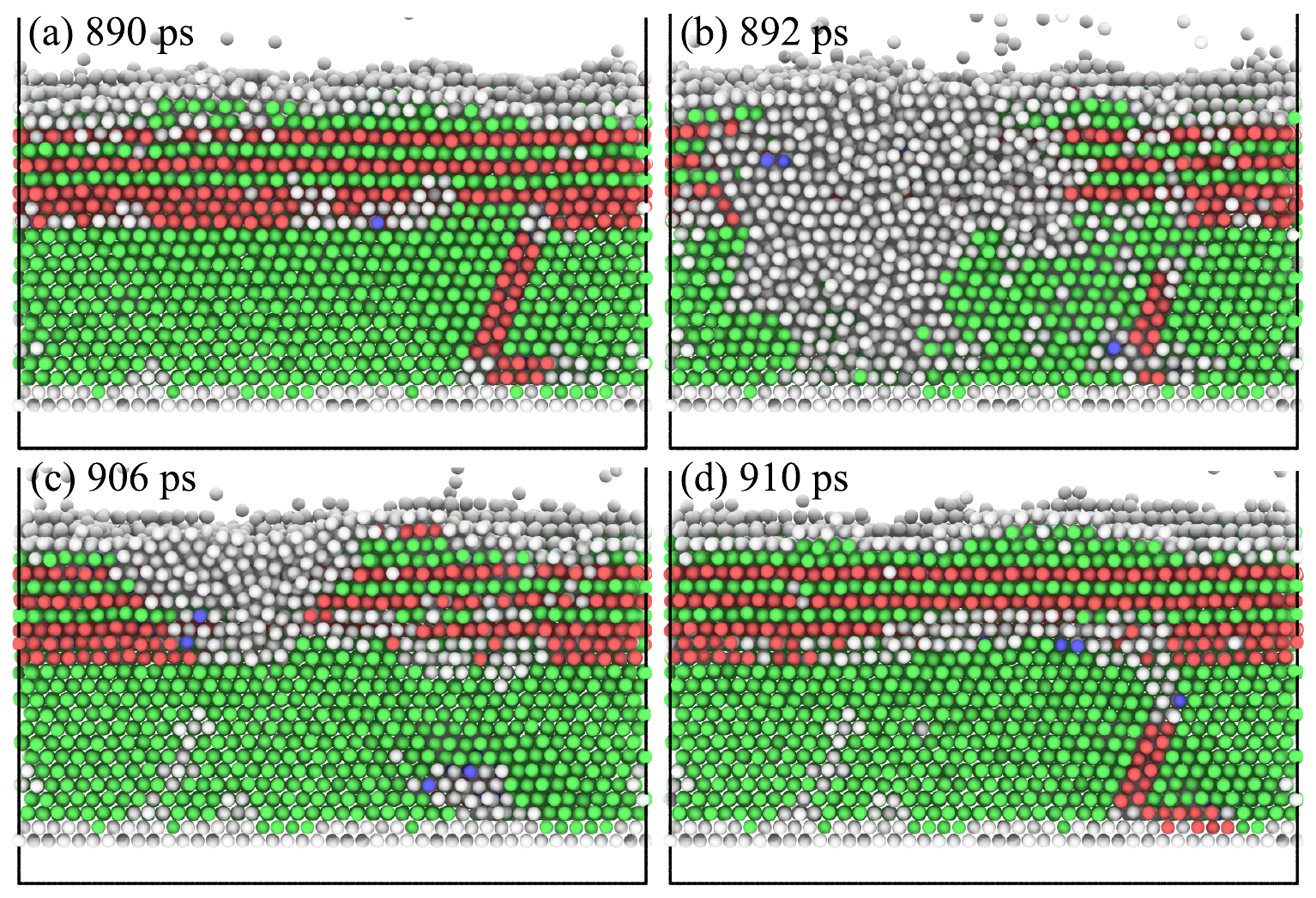}}
    \caption{The sequence of amorphization and crystallization during HiPIMS deposition due to high energy ion bombardment. (a) Before collision at 890~ps, (b) right after high energy collision at 892~ps, and (c -- d) after secondary collisions at 906 -- 910~ps. The red, green, blue and white atoms respectively are fcc, hcp, bcc and disordered atoms.}
    \label{fig:collision}
\end{figure}

Fig.~\ref{fig:collision} shows the sequence of amorphization and crystallization events during HiPIMS deposition. Fig.~\ref{fig:collision}(a) shows the film before collision and which seems single crystalline aside from some stacking fault areas. In Fig.~\ref{fig:collision}(b) it can be clearly seen that an amorphous region appears in the film deep down to the bottom of the substrate. The amorphization during bombardment has been reported previously. \citep{dong1998,houska2014} As time passes the amorphous phase disappears as shown in Fig.~\ref{fig:collision}(c) -- (d). It is worth noting that after amorphization and recrystallization the film microstructure has remained nearly unchanged.

\section{Conclusion}

Using MD simulations, it is shown that HiPIMS deposition presents a smoother surface than less ionized deposition methods representing dcMS and thermal evaporation. It is shown that the surface roughness is the product of clustering in the vapor phase and island growth on the substrate surface. The former can be reduced by increase in ionized flux fraction as a consequence of repulsion of ions of the same polarity. However reducing island growth is more complex and it occurs through so-called ``bi-collision'' of high energy ions. First a high energy ion implants into sublayers and causes local amorphization which fills the gaps between islands. Secondary ion bombardment causes recrystallization and maintains a smooth surface. There is no high energy ion in the thermal evaporation which presents an extremely rough surface. However, during dcMS deposition the number of bi-collision events are rare as detected by thermal spikes in the film. As a result, the dcMS process presents an intermediate roughness between thermal evaporation and HiPIMS. In the HiPIMS, fully ionized flux increases the number of high energy ions significantly and the probability of bi-collision events and thus minimum surface roughness is achieved. This also contributes to interface mixing and gives superior adhesion in HiPIMS deposition compared to other methods.  

\section*{Acknowledgments}
This work was partially supported by the University of Iceland Research Funds for Doctoral students, the Icelandic Research Fund Grant Nos.~196141, 130029 and 120002023 and the Swedish Government Agency for Innovation Systems (VINNOVA) contract No. 2014-04876.

\bibliography{ref}
\end{document}